# A Secure Hash Function MD-192 With Modified Message Expansion

Harshvardhan Tiwari
Student, CSE Department
JIIT
Noida, India
tiwari.harshvardhan@gmail.com

Dr. Krishna Asawa
Asst. Prof., CSE/IT Department
JIIT
Noida, India
krishna.asawa@jiit.ac.in

*Abstract*—Cryptographic hash functions play a central role in cryptography. Hash functions were introduced in cryptology to provide message integrity and authentication. MD5, SHA1 and RIPEMD are among the most commonly used message digest algorithm. Recently proposed attacks on well known and widely used hash functions motivate a design of new stronger hash function. In this paper a new approach is presented that produces 192 bit message digest and uses a modified message expansion mechanism which generates more bit difference in each working variable to make the algorithm more secure. This hash function is collision resistant and assures a good compression and preimage resistance.

## I. INTRODUCTION

Function of hash algorithms is to convert arbitrary length data into fixed length data hash value and they are used in cryptographic operations such as integrity checking and user authentication. For the cryptographic hash function following properties are required:

- Preimage resistance: It is computationally infeasible to find any input which hashes to any prespecified output.

- Second preimage resistance: It is computationally infeasible to find any second input which has the same output as any specified input.

- Collision resistance: It is computationally infeasible to find a collision, i.e. two distinct inputs that hash to the same result.

For an ideal hash function with an m-bit output, finding a preimage or a second preimage requires about $2^m$ operations and the fastest way to find a collision is a birthday attack which needs approximately $2^{m/2}$ operations [1].

The three SHA (Secure Hash Algorithms) algorithms [2, 7] SHA-0, SHA-1 and SHA-2 have different structures. The SHA-2 family uses an identical algorithm with a variable digest size. In the past few years, there have been significant research advances in the analysis of hash functions and it was shown that none of the hash algorithm is secure enough for critical purposes. The structure of proposed hash function, MD-192, is based on SHA-1. There are six chaining variables in suggested hash function. The extra 32 bit chaining variable makes the algorithm more secure against the brute force attack. The randomness of the bits in the working variables is not more when the original SHA-0 and SHA-1 codes were considered, because of this both SHA-0 and SHA-1 are totally broken using the differential attack by Wang[3,5,6]. Wang attacked on the poor message expansion of the hash function's compression function. In the suggested hash function a modified expansion mechanism is used, based on the modification to the standard SHA-1 hash function's message expansion proposed by Jutla and Patthak [11], in such a way that the minimum distance between the similar words is greater compared with SHA-0 and SHA-1. Because of the additional conditions in between the steps 16 and 79 there will be an additional security against the differential attack. Some other changes like, shifting of variables and addition of variables, have been made in order to make the algorithm more secure. The design goal of this algorithm is that, it should have performance as competitive as that of SHA-2 family.

## II. PREVIOUS WORKS

In this section we discuss about SHA hash functions and their weaknesses. The original design of the hash function SHA was designed by NSA (National Security Agency) and published by NIST in 1993. It was withdrawn in 1995 and replaced by SHA-1. Both SHA-0 and SHA-1 are based on the principle of MD5 [4] and are mainly used in digital signature schemes. They hash onto 160 bits and use Merkle-Damgard construction [1] from 160 x 512 → 160 compression function. At CRYPTO'98 Chabaud and Joux [9] proposed a theoretical attack on the full SHA-0 with the complexity of $2^{61}$. In 2004, Biham and Chen [10] presented an algorithm to produce near collisions. In 2005 Biham et al. presented optimization to the attack but the main improvement came from Wang. Both these algorithm (SHA-0 and SHA-1) generate a message digest of length 160 bits by accepting a message of maximum length $2^{64} - 1$ bits. In each of these hash function, message M is divided into r-blocks each of length 512bits such that, M= ($m_1$, $m_2$, $m_3$.........$m_r$).Then each block is further divided into sixteen 32 bit words such that $m_i$= $w_1$, $w_2$..........$w_{16}$, for $1 \leq i \leq r$. These 32 bit words are then linearly expanded into eighty 32 bit words $w_t$:

$w_t = w_{t-3} \oplus w_{t-8} \oplus w_{t-14} \oplus w_{t-16}$, for $16 \leq t \leq 79$

the only difference is that the SHA-1 uses a single bitwise rotation in the message schedule in its compression function where as SHA-0 does not. Both hash functions use an update function for processing each message block. This update function consists of eighty steps divided into four rounds. A,B,C,D,E are five 32 bit registers used as buffer for updating the contents. For each of the eighty rounds the registers are updated with a new 32 bit value. The starting value of these registers is known as initial value represented as $IV_0 = (A_0, B_0, C_0, D_0, E_0)$. In general, $IV_t = (A_t, B_t, C_t, D_t, E_t)$ for $0 \leq t \leq 79$. For step t the value $w_t$ is used to update the whole registers. Each step uses a fixed constant $k_t$ and a bitwise Boolean operation F which depends on the specific round, IF B THEN C ELSE D in first round, B XOR C XOR D in second and fourth round, MAJ(B,C,D) in third round. The process can be formally represented as:

$(A_t, B_t, C_t, D_t, E_t) = ((w_{t-1}+ A_{t-1}<<5+F(B_{t-1}, C_{t-1}, D_{t-1})+ E_{t-1}+ k_{t-1}), A_{t-1}, (B_{t-1}<<30), C_{t-1}, D_{t-1})$

In 2002 NIST developed three new hash functions SHA-256,384 and 512 [2] whose hash value sizes are 256,384 and 512 bits respectively. These hash functions are standardized with SHA-1 as SHS(Secure Hash Standard),and a 224-bit hash function, SHA-224, based on SHA-256,was added to SHS in 2004 but moving to other members of the SHA family may not be a good solution, so efforts are underway to develop improved alternatives.

### III. DESCRIPTION OF MD-192

The new dedicated hash function is algorithmically similar to SHA-1. The word size and the number of rounds are same as that of SHA-1.In order to increase the security aspects of the algorithm the number of chaining variables is increased by one (six working variables) to give a message digest of length 192 bits. Also a different message expansion is used in such a way that, the message expansion becomes stronger by generating more bit difference in each chaining variable. The extended sixteen 32 bit into eighty 32 bit words are given as input to the round function and some changes have been done in shifting of bits in chaining variables. Steps of algorithm are as follows:

Step 1: Padding   The first step in MD-192 is to add padding bits to the original message. The aim of this step is to make the length of the original message equal to a value, which is 64 bits less than an exact multiple of 512. We pad message M with one bit equal to 1, followed by a variable number of zero bits.

Step 2: Append length   After padding bits are added, length of the original message is calculated and expressed as 64 bit value and 64bits are appended to the end of the original message + padding.

Step 3: Divide the input into 512bit blocks   Divide the input message into blocks, each of length 512bits, i.e. cut M into sequence of 512 bit blocks $M^1, M^2, \ldots M^N$ Each of $M^i$ parsed into sixteen 32bit words $M^i_0, M^i_1, \ldots M^i_{15}$.

Step 4: Initialize chaining variables   $H^0$ = IV, a fixed initial value. The hash is 192 bits used to hold the intermediate and final results. Hash can be represented as six 32 bit word registers, A,B,C,D,E,F. Initial values of these chaining variables are:

A = 01234567

B = 89ABCDEF

C = FEDCBA98

D = 76543210

E = C3D2E1F0

F = 1F83D9AB

The compression function maps 192 bit value H=(A,B,C,D,E,F) and 512 bit block $M^i$ into 192 bit value. The shifting of some of the chaining variables by 15 bits in each round will increase the randomness in bit change in the next successive routines. If the minimum distance of the similar words in the sequence is raised then the randomness will significantly raises. A different message expansion is employed in this hash function in such a way that the minimum distance between the similar words is greater compared with existing hash functions.

Step 5: Processing      After preprocessing is completed each message block is processed in order using following steps:

   I)      For i = 1 to N prepare the message schedule.

$$W_t = \begin{cases} M^i_t, & 0 \leq t \leq 15 \\ W_{t-3} \oplus W_{t-8} \oplus W_{t-14} \oplus W_{t-16} \oplus \\ ((W_{t-1} \oplus W_{t-2} \oplus W_{t-15}) <<< 1), & 16 \leq t < 20 \\ W_{t-3} \oplus W_{t-8} \oplus W_{t-14} \oplus W_{t-16} \oplus \\ ((W_{t-1} \oplus W_{t-2} \oplus W_{t-15} \oplus W_{t-20}) <<< 1), & 20 \leq t \leq 63 \\ W_{t-3} \oplus W_{t-8} \oplus W_{t-14} \oplus W_{t-16} \oplus \\ ((W_{t-1} \oplus W_{t-2} \oplus W_{t-15} \oplus W_{t-20}) <<< 13), & 64 \leq t \leq 79 \end{cases}$$

**Figure1. Expansion of Message words**

   II)     Initialize the six working variables A,B,C,D,E,F with (i-1)st hash value.

III)  For t = 0 to 79
{
$P = ROTL^5(A) + F1(B,C,D) + E + K_t + W_t$

$Q = ROTL^5(A) + F1(B,C,D) + E + F + K_t + W_t$

$F = P$

$E = ROTL^{15}(D)$

$D = C$

$C = ROTL^{30}(B)$

$B = A$

$A = Q$
}

Where $K_t$ is a constant defined by a Table 1, F1 is a bitwise Boolean function, for different rounds defined by,

$F1(B,C,D) = IF\ B\ THEN\ C\ ELSE\ D$

$F1(B,C,D) = B\ XOR\ C\ XOR\ D$

$F1(B,C,D) = MAJORITY(B,C,D)$

$F1(B,C,D) = B\ XOR\ C\ XOR\ D$

Where the " IF….THEN……ELSE " function is defined by

$IF\ B\ THEN\ C\ ELSE\ D = (B \wedge C) \vee ((\neg B) \wedge D)$

and " MAJORITY " function is defined by

$MAJ(B,C,D) = (B \wedge C) \vee (C \wedge D) \vee (D \wedge B)$

Also, ROTL is the bit wise rotation to the left by a number of positions specified as a superscript.

IV)  $H_0^{(i)} = A + H_0^{(i-1)}$

$H_1^{(i)} = B + H_1^{(i-1)}$

$H_2^{(i)} = C + H_2^{(i-1)}$

$H_3^{(i)} = D + H_3^{(i-1)}$

$H_4^{(i)} = E + H_4^{(i-1)}$

$H_5^{(i)} = F + H_5^{(i-1)}$

| Rounds | Steps | F1 | $K_t$ |
|---|---|---|---|
| 1 | 0-19 | IF | 5a827999 |
| 2 | 20-39 | XOR | 6ed6eba1 |
| 3 | 40-59 | MAJ | 8fabbcdc |
| 4 | 60-79 | XOR | ca62c1d6 |

**Table1. Coefficients of each round in algorithm**

| Function | SHA-1 | SHA-256 | MD-192 |
|---|---|---|---|
| Block length (bits) | 512 | 512 | 512 |
| Message Digest Length (bits) | 160 | 256 | 192 |
| Rounds | 80 | 64 | 80 |
| Collision complexity (bits) | $2^{80}$ | $2^{128}$ | $2^{96}$ |

**Table2. Comparison among SHA-1, SHA-256 and MD-192**

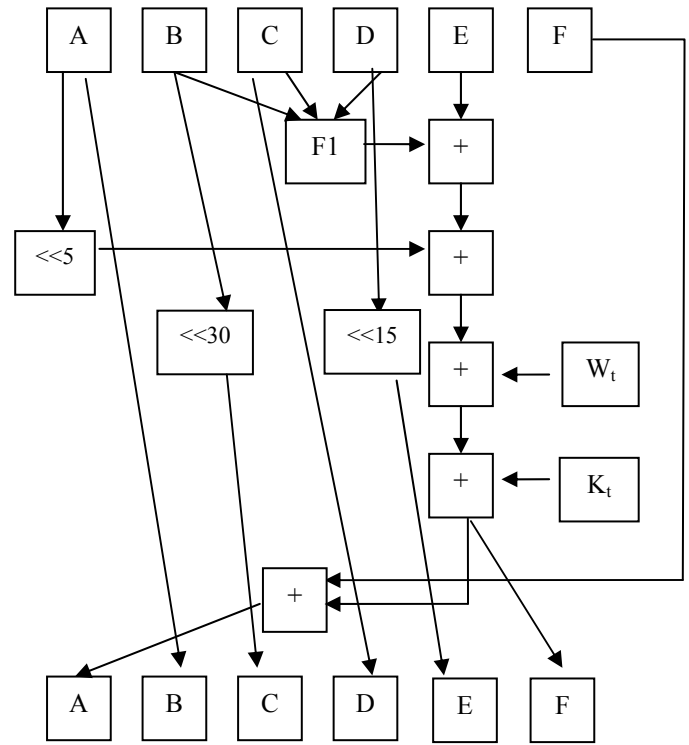

**Figure2. Proposed MD-192 step function**

IV.  PERFORMANCE

We have presented a new dedicated hash function based on Davies-Meyer scheme that satisfied Merkle-Damgard condition. Security of this algorithm is higher than SHA-1.Sophesticated message modification techniques were applied. This scheme is 192 bits and need $2^{96}$ bits for birthday paradox and is strong enough to preimage and second preimage attack. The performance of MD-192 is compared with SHA-1. The performance comparison is accomplished using Pentium IV, 2.8 GHz, 512MB RAM/ Microsoft Windows XP Professional v.2002. Simulation results of text data indicate that suggested algorithm needs more time to generate a message digest when compared with SHA-1 because in proposed algorithm there is an extra 32 bit chaining variable and additional conditions in between the steps 16 and 79 in message expansion mechanism. It produces message digest of length 192 bits longer than the SHA-1. From the simulation results of text data we have analyzed that strength of MD-192 is more than SHA-1. Even with the small change in the input algorithm produces greater change in the output.

## V. CONCLUSION AND FUTURE WORK

In this paper We proposed a new message digest algorithm basis on the previous algorithm that can be used in any message integrity or signing application. Future work can be made on this to optimize time delay.

### REFERENCES

[1] Ilya Mironov, "Hash Functions : Theory, attacks and applications", (Pub Nov 2005) J. Clerk Maxwell, A Treatise on Electricity and Magnetism, 3rd ed., vol. 2. Oxford: Clarendon, 1892, pp.68–73.

[2] NIST, "Secure Hash Standars",FIPS PUB 180-2,(Pub Aug 2002)

[3] X. Wang, X. D. Feng, X. Lai and H.Yu, "Collisions for Hash Functions MD4, MD5, HAVAL-128 and RIPEMD, (Pub Aug 2004) Available: http://eprint.iacr.org/2004/199/

[4] R.L. Rivest. The MD5 Message Digest Algorithm. RFC 1321, 1992

[5] X. Wang, H. Yu and Y.L. Yin, "Efficient Colision Search Attacks on SHA-0",(Pub 2005)

[6] K. Matusiewicz and J. Pieprzyk, "Finding good differential patterns attacks on SHA-1", (Pub 2004),Available: http://eprint.iacr.org/2004/364.pdf

[7] NIST, "Secure Hash Standar",FIPS PUB 180-1,(Pub Apr 1995)

[8] William Stallings, "Cryptography and Network Security: Principles and Practice. Third edition, Prentice Hall.2003.

[9] Florent Chabaud, Antoine Joux, "Differential collisions in SHA-0," Advances in Cryptology-CRYPTO'98, LNCS 1462, Springer-Verlag, 1998.

[10] Eli Biham, Rafi Chen, Antoine Joux, Patrick Carribault, Christophe Lemuet, William Jalby, "Collision in SHA-0 and Reduced SHA-1," Advances in Cryptology-EUROCRYPT 2005, LNCS 3494, Springer-Verlag,2005.

[11] C.S. Jutla and A.C.Patthak, "Provably Good Codes for Hash Function Dessign, (Pub Jan 2009)

| Message | SHA-1 | MD-192 |
|---|---|---|
| " " | da39a3ee5e6b4b0d3255bfef95601890afd80709 | 0fadadefc0ef131b93aa5854a29a0b506769fd32a6c90def |
| "a" | 86f7e437faa5a7fce15d1ddcb9eaeaea377667b8 | 4bd559a131498fcf07d06b2bf6ab8c4ccff1f5b3c4dce3c8 |
| "abc" | a9993e364706816aba3e25717850c26c9cd0d89d | b6a3a4d1a96e22d795c4f6db7d72607eea6d72fb7a440960 |
| "ABCDEFGHIJKLMNOPQRSTUVWXYZ" | 80256f39a9d308650ac90d9be9a72a9562454574 | 69791d6198d7d65d264e5f39a2bd426a341eb5dfd3aec5a8 |
| "abcdefghijklmnopqrstuvwxyz" | 32d10c7b8cf96570ca04ce37f2a19d84240d3a89 | 86c4ef2b05f8080bb041635aae7e0c60cf17bf1a6254ae8d |
| "a1b2c3d4e5f6g7h8i9j10" | df7175ff3caef476c05c9bf0648e186ea119cce7 | 034c641bb987efd91c6a73221c9da9ded649fddfa0986905 |
| "A1B2C3D4E5F6G7H8I9J10" | 28b083ed69254a8304f287aefe8d91295625beb0 | 76c6867583b9e4efaa6bdd350f6d527031c567db5a557a32 |
| "102030405060708090100100908070605040302010109876543211234567891 0" | 2604f26a461885848f54ce3b411bac69c31c140d | 5677b63d33afb99963e98e6d9705d49f327b90e7ca2e1216 |

**Table3. Message digest for certain messages**